# AN X-RAY SELECTED GALAXY CLUSTER IN THE LOCKMAN HOLE AT REDSHIFT 1.753

J. Patrick Henry[1, 2], Mara Salvato[3], Alexis Finoguenov[2, 4], Nicolas Bouche[5], Hermann Brunner[2], Vadim Burwitz[2], Peter Buschkamp[2], Eiichi Egami[6], Natasha Förster-Schreiber[2], Sotiria Fotopoulou[7], Reinhard Genzel[2], Günther Hasinger[3], Vincenzo Mainieri[8], Manolis Rovilos[2], Gyula Szokoly[2, 9]

[1] Institute for Astronomy, University of Hawaii, 2680 Woodlawn Drive, Honolulu, HI 96822, USA
[2] Max Planck Institut für extraterrestrische Physik, Giessenbachstr., 85748 Garching, Germany
[3] Max Planck Institut für Plasmaphysik and Excellence Cluster, Boltzmannstr. 2, 85748 Garching, Germany
[4] University of Maryland Baltimore County, 1000 Hilltop Cr., Baltimore, MD, 21250, USA
[5] Department of Physics, University of California, Santa Barbara, Santa Barbara, CA 93106, USA
[6] Steward Observatory, University of Arizona, 933 N. Cherry Ave., Tucson, AZ 85721, USA
[7] Technische Universität München, James-Frank-Str., 85749 Garching, Germany
[8] ESO, Karl-Schwarschildstr. 2, 85748 Garching, Germany
[9] Institute of Physics, Eötvös University, Pázmány P. s. 1/A, 1117 Budapest, Hungary

ABSTRACT

We have discovered an X-ray selected galaxy cluster with a spectroscopic redshift of 1.753. The redshift is of the brightest cluster galaxy (BCG), which is coincident with the peak of the X-ray surface brightness. We also have concordant photometric redshifts for seven additional candidate cluster members. The X-ray luminosity of the cluster is $3.68 \pm 0.70 \times 10^{43}$ erg s$^{-1}$ in the 0.1 – 2.4 keV band. The optical/IR properties of the BCG imply its formation redshift was ~5 if its stars formed in a short burst. This result continues the trend from lower redshift in which the observed properties of BCGs are most simply explained by a single monolithic collapse at very high redshift instead of the theoretically preferred gradual hierarchical assembly at later times. However the models corresponding to different formation redshifts are more clearly separated as our observation epoch approaches the galaxy formation epoch. Although our infrared photometry is not deep enough to define a red sequence, we do identify a few galaxies at the cluster redshift that have the expected red sequence photometric properties.

*Subject headings*: galaxies: clusters: general – galaxies: clusters: individual (XMMU J105324.7+572348) – galaxies: formation – X-rays: general

## 1. INTRODUCTION

Galaxy clusters provide information on the formation and evolution of both the large-scale structure in which they are embedded and the galaxies they contain. The former can be used for cosmology studies, such as determining the amount and nature of dark energy, while the later provides information on the formation process of galaxies. The clusters that provide the most leverage for these studies are naturally those at the highest redshifts since they give the most direct view of the cluster and galaxy formation process. High redshift in this context is ~2, since the observed structures surrounding $z > 2$ radio galaxies are interpreted as being protoclusters (see the review by Miley & De Breuck, 2008).

Thus searching for distant clusters is currently a very active field. Finding clusters at z > 1.5 has however proven to be a difficult task. The most distant object with a spectroscopic redshift published to date is IRC 0218-A = SXDF-XCLJ0218-0510 at z = 1.62 (Papovitch et al., 2010; Tanaka, Finoguenov and Ueda, 2010). There are several published cluster candidates with photometric redshifts in the range 1.75 – 2.00 (Andreon et al. 2009, but see Bielby et al. 2010; Chiaberge et al. 2010). The dearth of known clusters at these redshifts comes from the significant observational challenges associated with all the previously successful X-ray, optical, IR and radio selection techniques, which include object faintness and foreground contamination. In addition, the number of clusters will decrease as observations approach the epoch at which we believe they formed (z ~ 2) making them even more difficult to find.

We report here the discovery and spectroscopic redshift of the X-ray selected galaxy cluster XMMU J105324.7+572348 (hereafter LH146) and discuss its properties. At z = 1.75, this object is the most distant cluster reported to date with a spectroscopic redshift. This object is an extremely faint X-ray source in the Lockman Hole and the elucidation of its properties is only possible from our multi-decade X-ray and ground-based follow up work. In addition to the cluster X-ray properties we also give some photometric and spectroscopic properties of the brightest cluster galaxy (BCG), which we define as the brightest object at K close to the X-ray surface brightness peak; the quantitative selection criteria are in Section 3.1. Unless noted, we use Vega magnitudes in this paper. We do not correct for extinction since the Lockman Hole has the minimum value on the sky (Hydrogen column density ~6 x $10^{19}$ $cm^{-2}$ or E(B-V) ~ 0.008). We assume a cosmology with parameters $H_0$ = 72 km $s^{-1}$ $Mpc^{-1}$, $\Omega_m$ = 0.25 and $\Omega_\Lambda$ = 0.75. For this cosmology the redshift of LH146 implies an angular diameter distance of 1736 Mpc or 0.505 Mpc $arcmin^{-1}$; the look back time to it is 9.9 Gyr or 3.9 Gyr since the Big Bang.

## 2. X-RAY DATA

Brunner et al. (2008) describe the X-ray data we used. Briefly, the effective on-axis XMM-Newton exposure after flare removal is ~470 ks per EPIC detector. The standard XMM-Newton source detection algorithm was used to find sources within 15′ of the field center in the co-added data from the three EPIC detectors. Objects with a detection likelihood in the 0.5 – 10.0 keV band ≥ 10 were retained. The source under consideration here is number 146 in Table 3 of Brunner et al. (2008), 7.5′ off axis with a detection likelihood of 10 and with a 0 likelihood of being extended in their analysis. Monte Carlo simulations show that at off-axis angles ≤ 7.5′ there would be 0.1 spurious sources at this significance. There is another source, number 137 in Brunner's Table 3 (hereafter LH137), that is only 25" distant from LH146, but whose 0.5 – 2.0 keV flux is 5.5 times brighter in their analysis and also found to be a point source. Most high latitude X-ray point sources are AGN. We have not been able to confirm this suspicion for LH137 due to its faintness, z′ = 25.8 ± 0.3.

We show the exposure divided, background subtracted X-ray image around these two sources in Figure 1. LH146 appears diffuse to our eyes and it was this property that first drew our interest to it among the myriad of other faint Lockman Hole sources. The original fits reported in Brunner et al. (2008) were tuned to find point sources, so only used data within the 68% encircled energy radius (17" at the position of LH137 and LH146). Performing simultaneous fits to the two sources, but using counts within a 36" radius of each, increased the detection likelihood of LH146 from 10 to 23 (6.3σ equivalent Gaussian significance or 0.08 expected spurious sources) and increased the extent likelihood from 0 to 4.1 (2.1σ equivalent Gaussian significance). We find that LH137 remained a point source in this new fit with position and 0.5 – 2.0 keV flux largely unchanged ($10^h$ $53^m$ $27.41^s$ to $27.74^s$; +57° 23′ 35.9" to 35.4"; 12.2 ± 0.9 to 14.4 ± 0.9 x $10^{-16}$ erg $cm^{-2}$ $s^{-1}$). The detection likelihood is smaller than in the point source fit in Brunner et al. (2008) (216 to 186). The different outcomes from the two fits are expected if LH137 is a point source and LH146 is a diffuse source possibly containing a compact component that helped the original analysis find it.

We next present the X-ray spatial structure of LH146 using tools designed to characterize diffuse objects. In particular these tools do not assume a particular surface brightness distribution, as does the standard XMM-Newton analysis presented so far. The presence of LH137 considerably complicates this analysis since the separation of the two sources is the XMM – Newton mirror 80% encircled energy radius. That is, the wings of the PSF of LH137 extend to the position of LH146.

Adaptively smoothing the X-ray image using the procedure of Ebeling, White, & Rangarajan (2006) shows that indeed this situation pertains. However, we can compare the two sources if we plot contours at the same fraction of the peak surface brightness for each, which we show in Figure 2a. Although the absolute levels of the contours for the two sources differ, this display clearly shows that LH146 is more extended than LH137.

We have developed a set of procedures using wavelet transforms designed to detect and characterize extended X-ray sources and have applied them to several deep XMM-Newton fields (Finoguenov et al. 2007, 2009 & 2010). In addition to the standard processing of the EPIC data, done using XMMSAS 6.5, we performed a more conservative determination of data affected by solar flares as described in Zhang et al. (2004). For each EPIC chip separately we exclude times when the 10 – 15 keV light curve binned in 100 s intervals is > 2σ above its average. We determine this level iteratively using the count rate histogram below the threshold. We also checked for higher than average background that sometimes occurs in one or two MOS chips (Kuntz & Snowden 2008), finding none. Next we removed the background due to various instrumental and astrophysical sources using a procedure we term quadruple background subtraction because it involves four steps. First, we remove energy bands around the Al Kα line for the pn and both MOS detectors and the Si Kα line for just the MOS detectors. Second, we subtract out-of-time events for the pn detector. Third and fourth, we subtract the instrumental and sky backgrounds by normalizing the templates provided by Lumb et al. (2002, 2003). The normalization procedure takes advantage of the different spatial distributions of these two backgrounds across the face of the detector (unvignetted and vignetted). We iterate steps three and four, incorporating the PSF model image described below to exclude regions that contribute statistically to the background although their individual contributions are undetectable.

XMM-Newton reaches its confusion limit with exposures of ~100 ks, which is much less than our exposure. So subtracting the emission from point-like sources is mandatory for any analysis of extended sources such as clusters. We employed a wavelet transform analysis in the 0.5 to 2.0 keV observed band as described in Finoguenov et al. (2007 & 2010). We used the so-called *à trous* wavelets on five scales increasing by a factor of 2 from 8" to 128". A short summary of the properties of these wavelets is in the Appendix of Henry, Finoguenov & Briel (2004). The point sources are detected and characterized with the 8" and 16" scales. Then model point spread functions with the appropriate fluxes at the positions of all detected sources are subtracted from the image. The wavelet analysis then continues on this residual image at the 32", 64" and 128" scales, which detects the extended sources and characterizes their net count rate (hence flux, luminosity and related quantities) and image. We describe how we obtain these two outputs in the following.

The net count rate, $R_N$, is that within an aperture in the background and point source subtracted exposure divided image. The radius of the aperture used is derived from the wavelet filtering but the counts come from simple aperture photometry. We determine the cluster flux and luminosity within $r_{500}$ from $R_N$ using an iterative process. The $r_{500}$ radius, and the analogous $r_{200}$ discussed below, is the radius within the average cluster density is 500 or 200 times the critical density at the redshift of the cluster. The count rate-to-flux (0.5 – 2.0 keV) conversion and other parameters needed are computed initially for an APEC spectrum (Smith et al. 2001) for a temperature of 2 keV, 0.33 solar abundance at redshift 0.2, yielding 1.59 x $10^{-12}$ and 5.41 x $10^{-12}$ erg cm$^{-2}$ count$^{-1}$ for the pn and each MOS detectors respectively. The $r_{500}$ radius is 0.391 Mpc (kT)$^{0.63}$ E(z)$^{-1}$, where here and the following kT is in keV and E(z) = [$\Omega_m$ (1 + z)$^3$ + $\Omega_\Lambda$]$^{0.5}$. The aperture flux is then corrected (extrapolated) to $r_{500}$, denoted $f_{500}$, using a β-model with parameters β = 0.4 (kT)$^{0.333}$ and core radius = 0.07 (kT)$^{0.63}$ $r_{500}$. Next the luminosity in the 0.1 – 2.4 keV rest frame band is $L_{500}$ = 4π $D_L^2$ k(z, kT) $f_{500}$, where $D_L$ is the luminosity distance and k(z, kT) is the k correction now using the spectroscopic redshift. Finally, the revised kT comes from the luminosity – temperature relation: kT = 0.2 + 6.0 [$L_{500}$ E(z)$^{-1}$ (2.8 x $10^{44}$)$^{-1}$]$^{0.476}$ and the loop repeats until convergence. Leauthaud et al. (2010) determine that the mass within $r_{200}$ is $M_{200}$ = 5.37 x $10^{13}$ E(z)$^{-1}$ [$L_{500}$ E(z)$^{-1}$ (5.0 x $10^{42}$)$^{-1}$]$^{0.64}$. We note that except for mass-luminosity, these relations are well calibrated only for z < 0.3 and they could be different at the higher redshift of LH146.

Starting with net counts summed over the three EPIC detectors in the observed 0.5 – 2.0 keV band within a radius of 25″ of $R_N$ = 210.9 ± 40.4 and the equivalent MOS1 exposure of 1.883 Ms for the summed image, the above procedure yields $f_{500}$(0.5,2.0) = 6.37 ± 1.27 x $10^{-16}$ erg cm$^{-2}$ s$^{-1}$, making LH146 one of the faintest clusters ever observed (see the log N – log S plot in Rosati et al., 2002). The aperture correction to $r_{500}$ is a factor of 1.05. The luminosity in the rest frame 0.1 – 2.4 keV band is $L_{500}$(0.1,2.4) = 3.68 ± 0.70 x $10^{43}$ erg s$^{-1}$, implying a temperature of 1.7 keV and an $M_{200}$ = 4.5 x $10^{13}$ M☉. This mass implies $r_{200}$ = 0.396 Mpc.

The Brunner at al. flux from their Table 3 is f(0.5,2.0) = 2.2 ± 0.5 x $10^{-16}$ erg cm$^{-2}$ s$^{-1}$, implying that most of the source flux is not in the compact component shown in Figure 2a. We want to combine the adaptively smoothed and wavelet filtered images to produce a higher fidelity representation of the LH146 surface brightness. To that end we extracted a 19″ x 27″ sub-image from the background subtracted adaptively smoothed image centered on the peak surface brightness of LH146 and added it to the image formed from the two largest wavelets scales. We show the resulting X-ray contours in Figure 2b. We do not believe that the low surface brightness feature to the east of the peak of LH146 is unsubtracted residual emission from LH137 because we do not observe similar features near other point sources. Further, analysis identical to what we have done here of the XMM image of the Chandra Deep Field South field yields a 100% match with the extended sources detected in the Chandra image above a flux of 1 x $10^{-15}$ erg cm$^{-2}$ s$^{-1}$ (Giacconi et al., 2002). Of course point spread function issues on the scale of interest to us are negligible for the Chandra data. The significance of the extended emission interior to the lowest contour in Figure 2b is 3.8σ. This value includes estimates of the systematic errors from the point source and background subtractions. It is conservative in that it does not include the flux from the compact component near the X-ray surface brightness peak.

## 3. OPTICAL/IR DATA

### 3.1 Imaging

We give a summary of the majority of our imaging data in Table 1. Optical data come from two sources. We used the Large Binocular Camera (Giallongo et al. 2008), a prime focus imager on the 2 x 8.4 m Large Binocular Telescope, to obtain images in the $U_{LBC}$, $B_J$, $V_J$ and z′ bands. A description of the $U_{LBC}$, $B_J$ and $V_J$ observations and their reduction is in Rovilos et al. (2009); the z′ band was observed and reduced similarly. Complementary observations come from the IfA Deep Survey summarized by Barris et al. (2004). We used the SuprimeCam prime focus imager (Miyazaki, et al. 2002) on the Subaru 8.3 m telescope to obtain images in the $R_c$, $I_c$, and z′ bands. Near infrared data come from the 3.8 m United Kingdom Infrared Telescope Infrared Deep Sky Survey (UKIDSS) described by Lawrence et al. (2007). The Lockman Hole is one of their Deep Extragalactic Survey fields, which have J and K images of LH146. We used Data Release 7plus. Mid infrared data at 3.6 μm, 4.5 μm, 5.8 μm and 8.0 μm come from the IRAC imager on the Spitzer Space Telescope. We denote these bands by IRAC1-4 respectively.

Our object list comes from the catalog presented in Fotopoulou et al. (2010; in preparation), which is the union of objects detected in the $R_c$ and z′ images. Magnitudes for every filter at the positions in the object list come from SExtractor running in dual mode. The coincidence radius for the Spitzer data was 1" in order to avoid incorrect matches due to the relatively large point spread function of the mid-infrared images. We work with aperture-corrected total magnitudes. For the optical data ($U_{LBC}$ through z′) we derive these aperture corrections by inserting fake stars into the actual data frames and then running the same analysis as for the objects of interest. For UKIDSS we obtain the total magnitudes directly from their data products. For the Spitzer data we use the aperture corrections in Table 9 of Surace et al. (2005). Fake star simulations also provide a measurement of the limiting magnitudes using the procedure described in Section 5.3 of Rovilos et al. (2009). We define our limiting magnitude as that at which we detect 50% of the fake stars. Table 1 lists these limits for the filters for which we performed these simulations.

Later we give some properties of the BCG, so describe here how it was chosen. We use the reddest band available (because small amounts of star formation can greatly brighten blue fluxes) that has the requisite depth and spatial resolution. This is the K band for the data available to us. In UKIDSS DR7p there are 8 objects with K brighter than 19.0, with ≥ 90% probability of being a galaxy and within a square centered on the X-ray peak with 2′ sides. The object we call the BCG is one of these and is ~2″ north of the X-ray peak, which is coincident with the peak within the astrometric errors. Six of the remaining seven objects are outside Figure 2. The last object is a bright galaxy at $\alpha(2000) = 10^h\ 53^m\ 21.4^s$, $\delta(2000) = +57°\ 23′\ 46.4″$. We do not consider this object to be a candidate for the BCG since it is outside the lowest X-ray contour in Figure 2b.

We determined photometric redshifts of objects in the Fotopoulou et al. (2010) catalog using the photoz code Le Phare (Arnouts, 2010) with the same procedure and library as used by Ilbert et al. (2009) and Salvato et al. (2009) in the COSMOS field. The training set consisted of a sample of 57 normal galaxies with reliable redshifts (Ingo Lehmann, PhD thesis). The training consisted of computing the average zero point offset for each photometric band from the best fit model spectrum at the known spectroscopic redshift. The zero point offset thus corrects small uncertainties in the photometry, due, for example, to the yet not well-tested response curves of LBC filters. The zero point offsets are usually very small (< 0.05 mag) and only for IRAC bands reach 0.4 mag. The IRAC correction is surprisingly large compared to what we found in other fields (e.g. Ilbert et al., 2009 and Salvato et al., 2009 report IRAC offsets < 0.17 for the COSMOS field). The IRAC zero point offsets depend on which bands are included, dropping to 0.2 if only IRAC1 and 2 are fit. This behavior may result from the small number of training galaxies with IRAC fluxes.

Our data set has some deficiencies compared to the near optimum case of the COSMOS or Chandra Deep Field South fields. The spectroscopic range of the training set is limited, only reaching to z = 0.7, which is substantially less than the value expected for LH146 (see Section 3.2). The best photometric redshift results are obtained when a large sample of training sources are used whose spectroscopic redshifts span the expected range of the entire sample. In addition, our IR data are relatively shallow. Photometric redshift accuracy quickly degrades for sources with large photometric uncertainties (as will be the case for objects near the flux limits of the images, see Ilbert et al., 2009 and Salvato et al., 2009 for a complete discussion) and with few photometric points long ward of the 4000 Å break. That said the photometric redshift of the BCG of 1.77 ± 0.06 is in very good agreement with the spectroscopic redshift of 1.753 ± 0.001. In Figure 3 we show the photoz fit for the BCG and seven other galaxies discussed in Section 4. All of these objects have photoz errors ≤ 0.1. Thus despite the deficiencies noted above, our procedure can yield quality photometric redshifts.

We also obtained on UT March 11, 2009 a 5,400 s exposure of the field through the $K_s$ filter using MOIRCS on the Subaru 8.3 m telescope (Ichikawa et al. 2006; Suzuki et al. 2008). This image is about the same depth as the UKIDSS K data, but the pixel size is nearly a factor of 3 smaller (0.117" on a side vs. 0.200"), enabling a more sensitive search for close companions as a direct comparison demonstrates. The observing sequence consisted of four nine-point dither patterns with slightly offset starting positions. Each dither pattern was approximately a circle of 15" diameter with three coadded 50 s exposures at each point. We reduced these data with a set of IDL scripts written by Wei-Hao Wang called SIMPLE (Simple Imaging and Mosaicing Pipeline; http://www3.asiaa.sinica.edu.tw/~whwang/). This pipeline ingests a set of dithered exposures and produces a background-subtracted, exposure-corrected, photometrically-calibrated (if a standard was observed) mosaic on a common astrometric system. We show the resulting image of the BCG in Figure 4. The seeing in this image is 0.85". There is a faint companion ~1" to the SE of the nucleus. We measure a combined $K_s$ = 18.69 ± 0.04 integrated within a radius of 4" (background annulus 5.15" – 8.07").

## 3.2 Spectroscopy

We used MOIRCS in multi-object spectroscopy mode mounted on the Subaru 8.3 m telescope. We observed on UT February 6, 2007 for an exposure of 16,800 s and on UT March 11, 2009 for an exposure of 9,000 s, yielding a total exposure of 25,800 s. Individual integrations of 450 s, 600 s, or 900 s depending on airmass were added together to form the final exposures. We used the zJ500 grating with 0.8" slits, yielding a resolution of ~500. The seeing was 0.4" – 0.8" and 0.8", respectively on the two nights. The dither was ±0.5" along the slitlets.

We reduced the data with MOIRCSMOSRED, a set of IDL scripts originally written by Youichi Ohyama and revised by Tian-Tian Yuan. The two MOIRCS chips were reduced separately. The reduction consisted of forming median dark and flat field frames. The only use of the median dark was to find bad pixels to construct the bad pixel mask; the dark signal was removed during sky subtraction. The sky subtraction of the bad-pixel masked individual integrations is the first step and it occurs before any manipulation of the raw images. For each individual integration we subtracted the average of the preceding and succeeding dithered integrations. The first (A) dither used only the succeeding B dither, similarly the last (B) dither used only the preceding A dither. We used a relatively bright reference star to measure the flexure across the dispersion and sky lines to measure flexure along the dispersion for each individual integration. Flexures between successive individual integrations were generally smaller than one pixel. We applied the flexure correction quantized in single pixel steps when needed. We also used the reference star to measure the average flux detected in an individual integration, which changes due to transparency and centering variations. These flux measurements were used as weights when summing the A and B dithers. The final reduced image is a 2D spectrum of counts vs. pixel comprised of the weighted sum of the sky-subtracted flexure-corrected A dithers minus the corresponding quantity of the B dithers.

Only at this stage did we transform the image. We traced the edge of the slits on the median flat to make a small rotation of the spectra so they were aligned along the rows of the image. The images were then flattened. We determined the conversion from pixel to wavelength from nine bright night sky lines and rebinned the image to uniform wavelength pixels. HIP 61534 was used to measure the system response needed to give a 2D spectrum of relative flux vs. wavelength. The absolute flux of the BCG comes from its measured Petrosian J magnitude of 20.94 ± 0.15 from the UKIDSS or $F_\lambda(1.24\ \mu m) = 1.32 \pm 0.18 \times 10^{-18}$ erg cm$^{-1}$ s$^{-1}$ Å$^{-1}$ using the calibration of Cohen, Wheaton & Megreath (2003). We then extracted a 1D spectrum of absolute flux vs. wavelength for the three different individual integrations and added them together, after making small adjustments to the wavelength and flux scales of the 2009 data so it is compatible with that of 2007. We show our final spectrum of the BCG in Figure 5. The measured redshift of the CaII H, K lines and the 4000 Å break is 1.752, 1.752, and 1.755, respectively, implying $z = 1.753 \pm 0.001$. By placing fake lines in the measured spectrum, we can place an upper limit on the [OII] $\lambda$3727 Å flux of $2 \times 10^{-17}$ erg cm$^{-2}$ s$^{-1}$. This limit is not in a formal statistical sense; rather it is the flux we would judge to be a real line. Using equation (3) of Kennicutt (1998) this limit implies a limit on the BCG star formation rate of 6 M$\odot$ per year. We placed slitlets on other objects in the field of our MOIRCS observation but could only measure a redshift for this one.

## 4. NATURE OF LH146

Specifically we want to answer the question whether LH146 is a cluster at $z = 1.75$. LH146 is a high galactic latitude X-ray source detected at the 6.3$\sigma$ confidence level whose spatial distribution is extended at the 3.8$\sigma$ confidence level. There are three known ways that a high galactic latitude X-ray source can be extended. These are source confusion; the source is a radio galaxy (Finoguenov et al., 2010) or a cluster of galaxies. We examine each of these possibilities in turn.

The limiting flux for the detection of a point source with the wavelet analysis is $1.4 \times 10^{-16}$ erg cm$^{-2}$ s$^{-1}$. Thus 4.6 such sources are needed to be confused in order to produce the observed flux. The log N – log S distribution derived at the Lockman Hole by Brunner et al. (2008) yields 2,764 sources deg$^{-2}$ at the flux limit or 0.43 expected sources within the solid angle subtended by LH146. The Poisson probability of having 4.6 sources when 0.43 are expected corresponds to approximately a 3.5$\sigma$ equivalent Gaussian fluctuation. This calculation is approximate because the Poisson distribution is only defined for integer values. The calculation is conservative because, with decreasing flux, the number of sources needed to be confused grows as $S^{-1}$ while the number available to be confused only increases as $S^{-0.55}$. We conclude that source confusion is an unlikely cause of the LH146 diffuse flux.

There is strong evidence that LH146 is not a radio galaxy. Biggs and Ivison (2006) present a very deep radio survey of the Lockman Hole and we use their data to place a 2$\sigma$ upper limit of 9.2 $\mu$Jy at 1.4 GHz on the flux of any source at the position of LH146. The nearest detected source is well outside the field of Figure 2. This limit corresponds to a luminosity of $2.1 \times 10^{23}$ W Hz$^{-1}$ at a rest frame frequency of 500 MHz, calculated according to the appendix of Miley & De Breuck (2008). This luminosity is well below their definition of a high-redshift radio galaxy.

By process of elimination then, LH146 is a cluster as there are no other known alternatives. Further its X-ray properties are similar to other low redshift clusters. An example is A262, which has L(0.5,2.0) = 2.45 x $10^{43}$ erg s$^{-1}$, and a temperature of 1.4 keV (David, Jones & Forman, 1996).

We have a single spectroscopic redshift of the galaxy with the brightest K magnitude within the X-ray contours and the projected location of this galaxy is consistent with the X-ray peak (Figure 2). The redshift of a single dominant galaxy under the peak of the X-ray surface brightness is invariably the redshift of the entire cluster. Hudson et al. (2010) find that 78% of cluster BCGs are within 12 kpc of the X-ray peak for local clusters (see also Lin & Mohr (2004) and Sanderson, Edge, & Smith (2009) for similar results). Collins et al. (2009, their Figure 1) give an example at z = 1.39. There are an additional seven objects with concordant photometric redshifts in a square region with 2′ = 1 Mpc sides centered on the BCG. The half width of this square is 1.3 $r_{200}$, i.e. the square encompasses the likely physical region of a cluster. We used a relatively strict definition of concordance: best fit photoz within ± 0.04 of the BCG spectroscopic value and a photoz error < 0.1. We list these objects in Table 2 in which the columns are a running identification number, the J2000 coordinates of the object, its derived photoz and 68% confidence error, and the magnitude and colors for the indicated bands. We show the photoz fits in Figure 3 and a finding chart for the objects in Figure 6. The only object in Table 2 that we observed with the MOIRCS spectrometer was the BCG, so we could not stack spectra to increase the signal to noise. Figure 7 shows the redshift histogram of all objects in the 2′ x 2′ square for which we have photozs and for which the photoz error is < 0.1. There is a peak at the BCG redshift that is clearly significant compared to the average photoz histogram for the same size solid angle away from LH146 (the Poisson probability of obtaining the observed 8 galaxies when 0.377 are expected is 6.94 x $10^{-9}$, equivalent to 5.7σ of a Gaussian distribution). Despite the significance, the galaxy over density is lower than some optical/IR selected very high redshift clusters (e.g. Papovitch et al., 2010), which may be expected when there are different selections (X-ray flux vs. galaxy over density for these two cases). Thus the totality of the spectroscopic and photometric data shows that the redshift of the cluster is 1.75.

## 5. DISCUSSION

We want to compare the J and $K_s$ magnitudes of the BCG in LH146 with those of lower redshift examples. We use the UKIDSS to obtain a Petrosian K magnitude = 18.62 ± 0.06 and an aperture 3 (1″ radius) color of J – K = 2.26 ± 0.08. We use this color because it is consistent with but has a smaller error than the Petrosian value of 2.32 ± 0.16. Hewett et al. (2006) find that $K_s$ = K + 0.068 and J – $K_s$ = J – K + 0.01. These imply Petrosian $K_s$ = 18.69 ± 0.06 and J – $K_s$ = 2.27 ± 0.08. The $K_s$ magnitude agrees very well with our independently measured value described in Section 3.1. We compare these values with lower redshift BCGs in Figure 8. The LH146 BCG has $z_f$ = 5, only 1.2 Gyr after the Big Bang and 2.7 Gyr prior to the epoch when we observe the cluster. Thus the photometric properties of the LH146 BCG are similar to other lower redshift BCGs. In general our data support and extend to earlier epochs the conclusions of Collins et al. (2009). Most (>90%) of the BCG mass is already in place by these redshifts, in conflict with many galaxy formation models, but some mass assembly is ongoing given the presence of close companions of the BCG that will likely merge with it.

There is one caveat to this conclusion. The LH146 BCG mass is within the range of masses of the z > 1.2 objects in Collins et al. (2009) using their scaling from $K_s$ absolute magnitudes. However the mass of the LH146 cluster is a factor of 2.5 to 6.6 smaller than the z > 1.2 clusters discussed by Collins et al. (2009), using the scaling between X-ray luminosity and cluster mass given in Section 2. The discrepancy is less severe (<15% smaller in the best case) if we compare mass fluctuations, the cluster mass divided by the background mass dispersion, which may be more physically relevant for objects at different redshifts. Nevertheless we should keep in mind that LH146 may not be identical to the lower z comparison objects in Collins et al. (2009).

Thus modulo this caveat, the dichotomy between the photometric data for BCGs, which resemble a single monolithic collapse, and the theoretically preferred gradual hierarchical formation process still remains. De Lucia & Blaizot (2007) resolve this dichotomy by proposing that the stars of all BCGs form very early (50% by $z \sim 5$) while the mass gradually assembles hierarchically much later (50% of the mass of a BCG destined to reside in a present day massive cluster resides in a single galaxy only after $z \sim 0.5$). Statistically, the BCG we observe in the highest X-ray luminosity clusters at $z = 1.75$ will not be the BCGs in the highest X-ray luminosity clusters at $z = 0$. Thus the observation of such high redshift BCGs does not violate the hierarchical assumption for most of them.

Whether the color-magnitude (CM) diagram of LH146 exhibits a red sequence (RS) similar to lower redshift clusters can test our understanding of the RS because we observe it within 3.9 Gyr of any star formation epoch, the age of the universe at the redshift of this cluster. The RS is believed to be a sequence of elliptical galaxy masses, with the more massive (brighter) galaxies retaining more metals (redder). The slope and width of the RSs observed to date show little evolution because all stars in ellipticals formed long before the observation epoch. But the colors of high and low metallicity ellipticals are significantly different for ages smaller than 4 Gyr leading to slope changes (Figure 3 of Gladders et al., 1998). Similarly, the fractional age (color) difference among galaxies of the same metallicity grow as the age (the denominator of the fractional age) decreases, implying a RS width at $z = 1.75$ that is three times larger than at $z = 0$ (Figure 11 of Gobat et al., 2008). Unfortunately our near IR photometry is too shallow to define any RS, only 2.2 magnitudes deeper in K than the BCG, particularly if the form of the RS is unknown. Nevertheless, there are three galaxies that are in the region of the CM diagram where the RS is expected to lie. That is their K magnitudes and z′-K colors in Table 2 are those of a population of stars that formed in a single burst at redshifts between 2.5 and 5. Figure 8 demonstrates some of these points in another way for one of those galaxies, the BCG. In addition, there is one other galaxy in Table 2 brighter in z′ than the BCG but not detected at J and K, indicating a different star formation history for it. Much deeper near IR photometric as well as spectroscopic observations are thus particularly interesting.

We thank the anonymous referee for comments that lead to an improved paper. JPH thanks the Alexander von Humboldt Foundation for support that enabled a visit to MPE while this paper was written. GH and MS acknowledge support from the German Deutsche Forschunggemeinschaft, DFG Leibniz Prize (FKZ HA 1850/28-1). NFS acknowledges support by the Minerva program of the MPG. This work is based in part on data collected at the Subaru telescope, which is operated by the National Observatory of Japan, XMM-Newton, an ESA science mission funded by contributions from ESA member states and from NASA, and the Spitzer Space Telescope, which is operated by the Jet Propulsion Laboratory, Caltech, under NASA contract 1407. We thank Ichi Tanaka and Kentaro Aoki for much help with the Subaru MOIRCS observations and Wei-Hao Wang and Tian-Tian Yuan for their help with reducing the data.
*Facilities*: Subaru (SuprimeCam, MOIRCS), LBT (LBC), UKIRT (UKIDSS), XMM-Newton, Spitzer (IRAC)

Table 1
Properties of the Data Used for Photometric Redshifts

| Filter | Telescope | Limit (AB)[1] | Exposure (s) | Seeing (″) |
|---|---|---|---|---|
| $U_{LBC}$ | LBT | 26.4 | 49,680 | 1.06 |
| $B_J$ | LBT | 26.6 | 19,972 | 0.90 |
| $V_J$ | LBT | 26.5 | 9,540 | 0.95 |
| $R_c$ | Subaru | 26.5 | 3,920 | 0.90 |
| $I_c$ | Subaru | 25.2 | 6,235 | 0.98 |
| $z'$ | Subaru | 25.4 | 10,640 | 0.96 |
| $z'$ | LBT | 24.0 | 14,400 | 1.06 |
| J | UKIRT | 23.4 | 8,960 | 0.84 |
| K | UKIRT | 22.9 | 13,740 | 0.73 |
| IRAC1 | Spitzer | 22.6 | 500 | 1.7 |
| IRAC2 | Spitzer | | 500 | 1.7 |
| IRAC3 | Spitzer | | 500 | 1.9 |
| IRAC4 | Spitzer | | 500 | 2.0 |

[1] 50% detection efficiency, total magnitude

Table 2
Properties of Candidate Cluster Members

| ID | α(2000) | δ(2000) | Photo z[1] | $z'$ | $z'-K$ | $z'-J$ |
|---|---|---|---|---|---|---|
| 88636 | 163.37520 | +57.38139 | $1.749^{+0.058}_{-0.050}$ | 22.35±0.02 | 3.40±0.04 | 1.53±0.07 |
| 89504 | 163.35770 | +57.38680 | $1.765^{+0.066}_{-0.060}$ | 22.89±0.03 | | |
| 91517 | 163.32703 | +57.39917 | $1.739^{+0.091}_{-0.063}$ | 24.29±0.10 | | |
| 91865 | 163.34630 | +57.40142 | $1.777^{+0.073}_{-0.089}$ | 24.25±0.10 | | |
| 91885 | 163.35210 | +57.40129 | $1.793^{+0.058}_{-0.056}$ | 24.49±0.12 | | |
| 171765[2] | 163.35243 | +57.39704 | $1.769^{+0.055}_{-0.061}$ | 23.44±0.05 | 4.65±0.06 | 2.34±0.09 |
| 182204 | 163.33470 | +57.41112 | $1.725^{+0.079}_{-0.088}$ | 24.64±0.14 | | |
| 202417 | 163.34364 | +57.38400 | $1.730^{+0.077}_{-0.066}$ | 22.98±0.03 | 4.03±0.05 | 1.77±0.09 |

[1] 1σ errors
[2] BCG with spectroscopic redshift = 1.753 ± 0.001

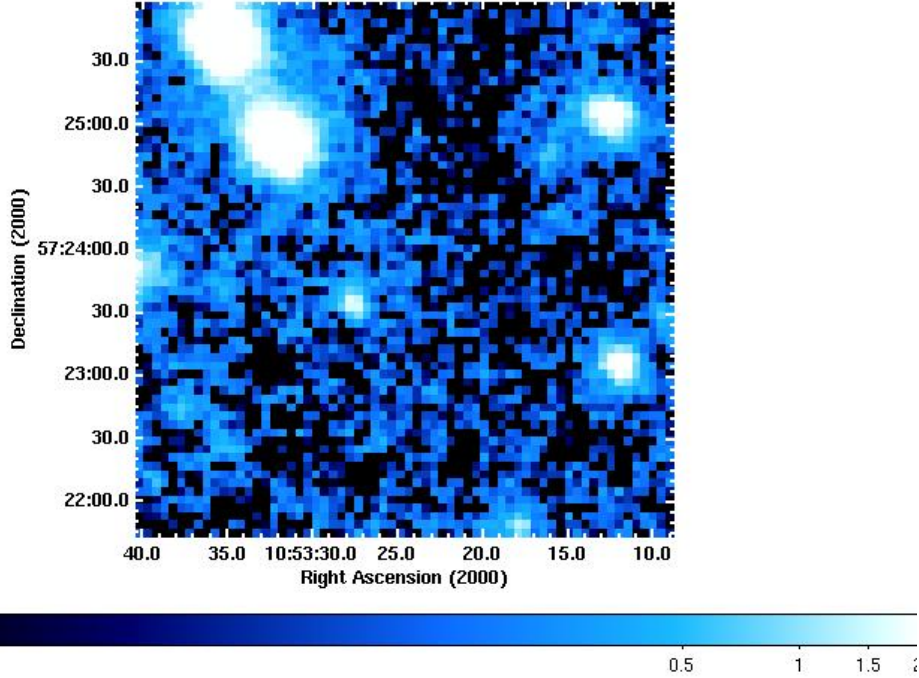

Figure 1. Exposure divided, background subtracted X-ray image in the 0.5 – 2.0 keV band centered on LH146 at $\alpha(2000) = 10^h\ 53^m\ 24.8^s$, $\delta\ (2000) = +57°\ 23'\ 48.2''$. LH137 is the source 25″ to the southeast of LH146. The color bar extends from 0.005 to 2.0 counts $(10^5\ s)^{-1}\ (4''\ x\ 4'')^{-1}$.

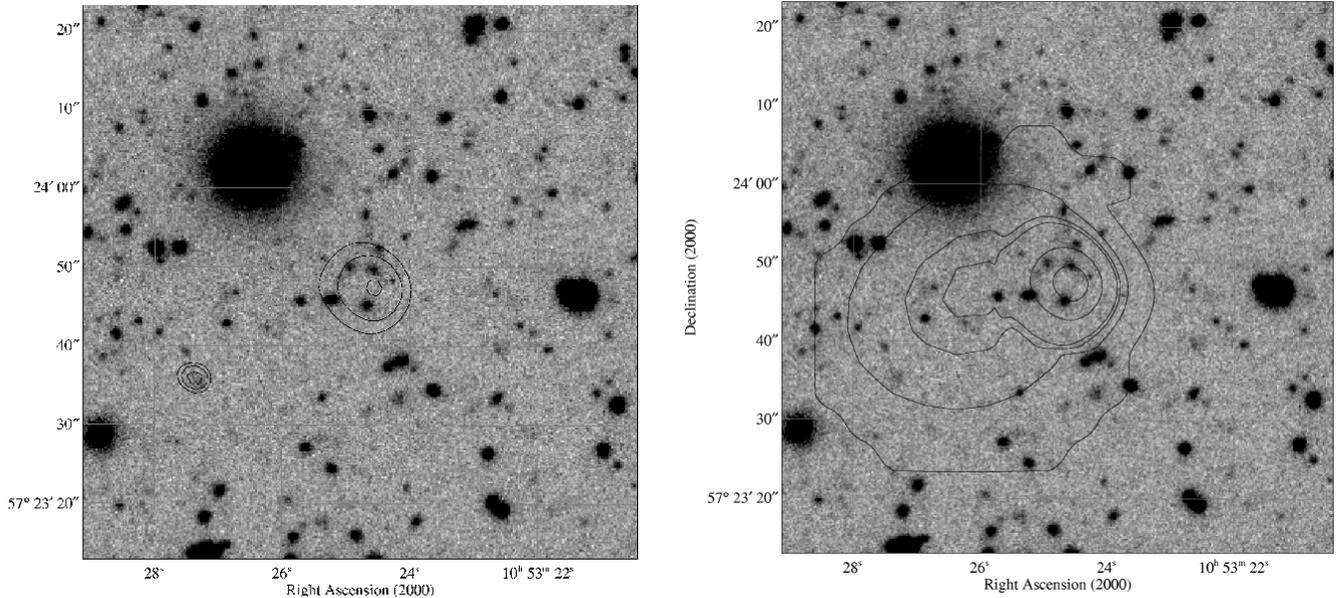

Figure 2. X-ray surface brightness contours overlaid on a Subaru SuprimeCam $I_c$ image. LH146 is at the center and LH137 is the point source to the southeast. The BCG, selected as described in Section 3.1, is immediately north of the peak of LH146. a. (left) Adaptively smoothed contours at 0.80, 0.89 and 1.00 times the peak intensity of both sources, that is they are at different absolute contours since the central surface brightness of LH137 is 54 times higher than LH146. b (right) Contours from an image composed of the 2 largest wavelet scales + a part of the adaptively smoothed LH146 image. The contours are at 1.2, 3.0, 3.5, 3.7, 4.4 and 4.9 x $10^{-16}$ erg cm$^{-2}$ s$^{-1}$ arcmin$^{-2}$ above background in the 0.5 – 2.0 keV band.

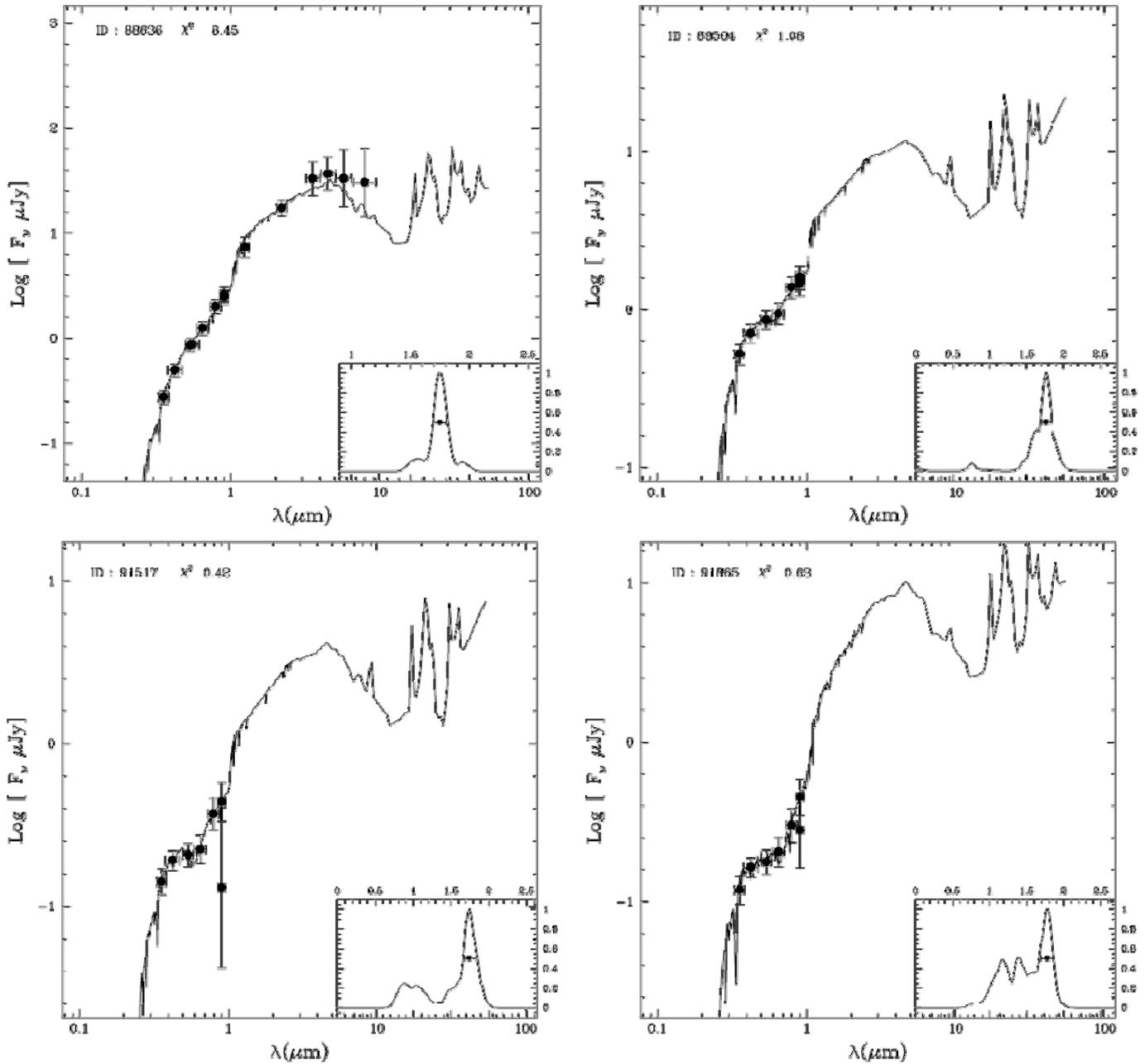

Figure 3. The observed spectral energy distributions of eight galaxies in a 2′ x 2′ square region around LH146 with the best fit galaxy template spectrum overlaid. The inset shows the redshift probability distribution. ID 171765 is the BCG. All these objects have a best fit photoz within ± 0.04 of the BCG spectroscopic redshift and photoz error < 0.1.

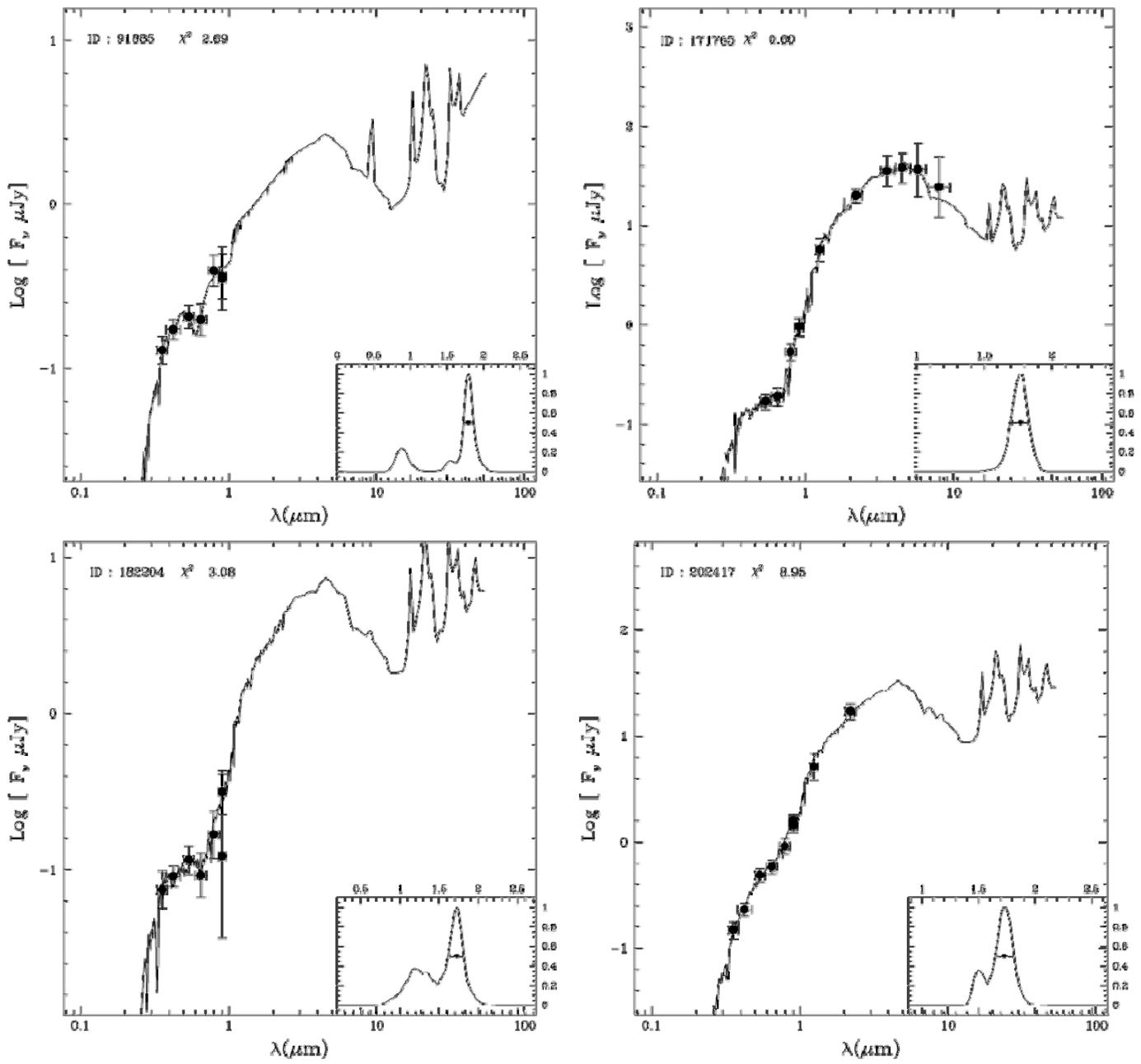

Figure 3. Continued.

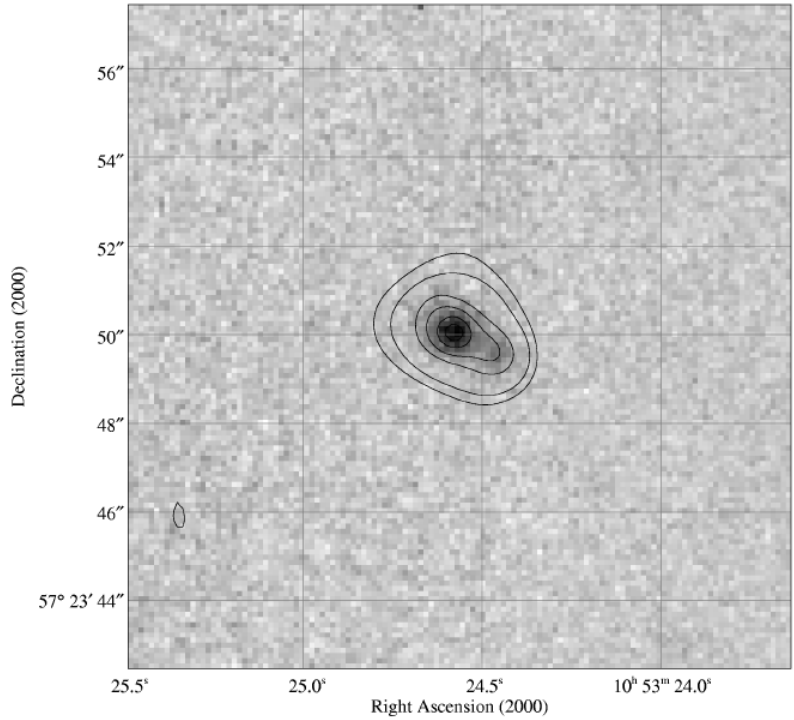

Figure 4. K$_s$ image of the BCG. The contours are from an adaptively smoothed image of the same data and are at 0.003, 0.006, 0.025, 0.05, 0.10, 0.15 µJy (0.117")$^{-2}$ above background.

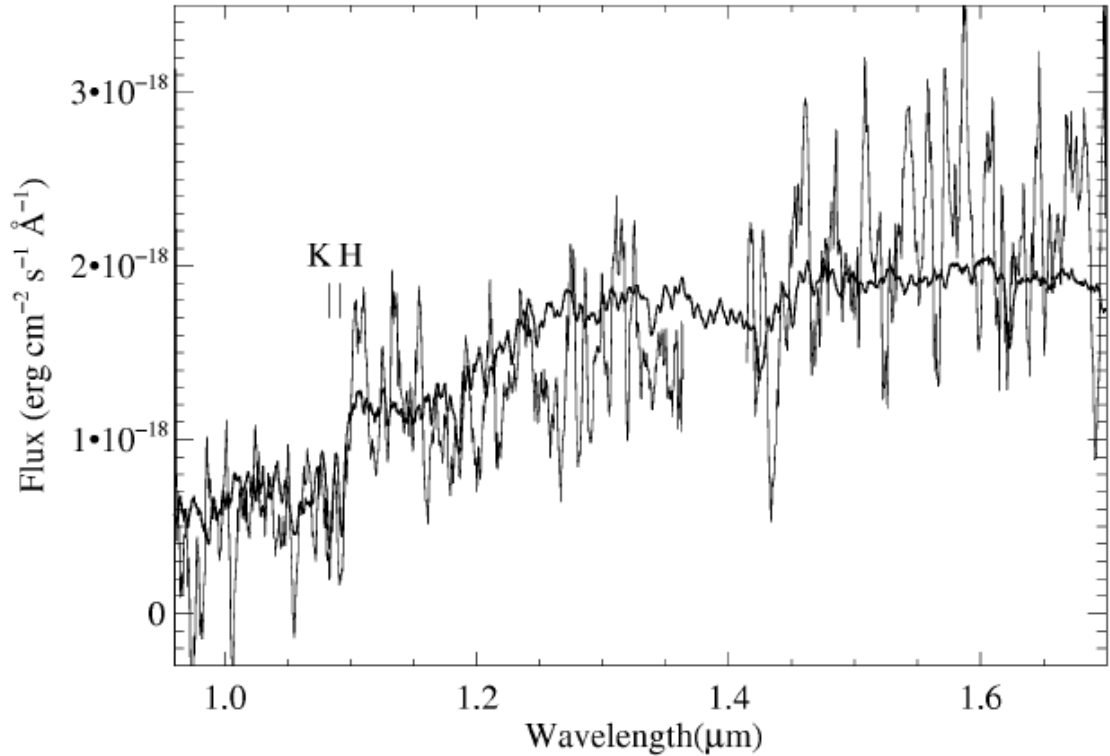

Figure 5. MOIRCS spectrum of the BCG with 50 Å boxcar smoothing (thin line) and the average elliptical spectrum from Eisenstein et al. (2003, Figure 4; thick line) redshifted to the cluster value of 1.753 overlaid. We do not show the wavelength interval from 1.365 µm – 1.415 µm due to its high noise caused by reduced atmospheric transparency.

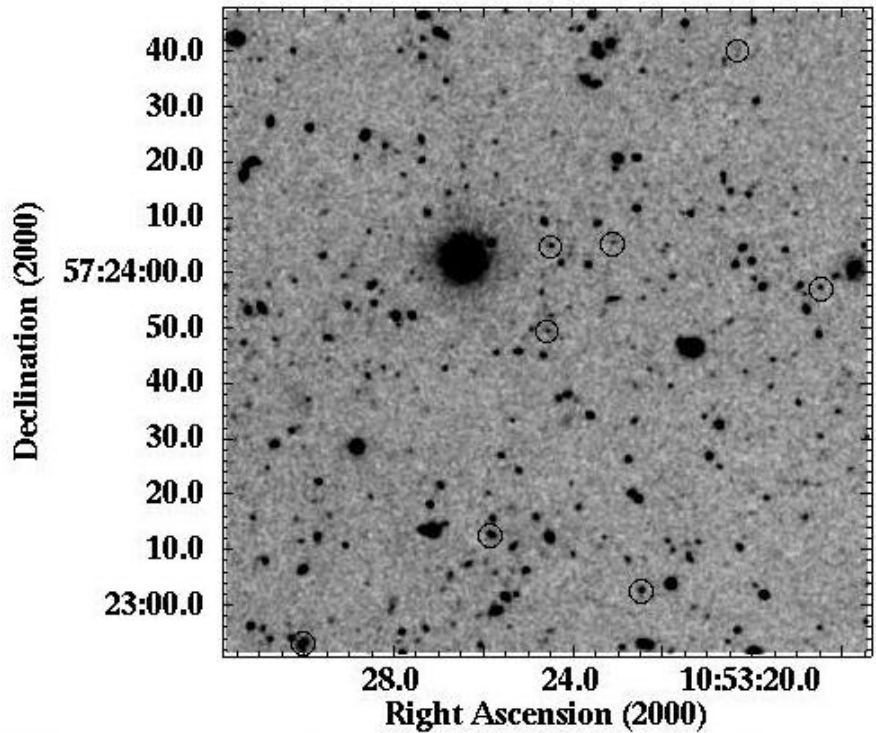

Figure 6. Finding chart of candidate cluster members in Table 2 using the Subaru SuprimeCam $I_c$ image.

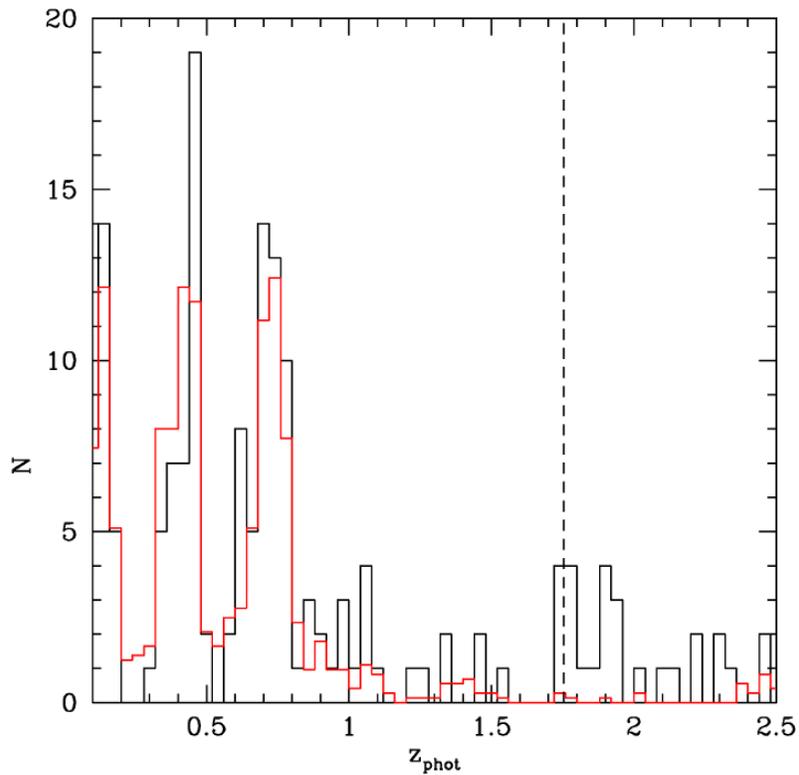

Figure 7. Photoz histogram for objects within a square of 2′ sides and with photoz error < 0.1. The black histogram is for a square centered on the BCG while the red histogram is the mean distribution in eight squares at random locations away from the cluster. The vertical dashed line is the spectroscopic redshift of the BCG.

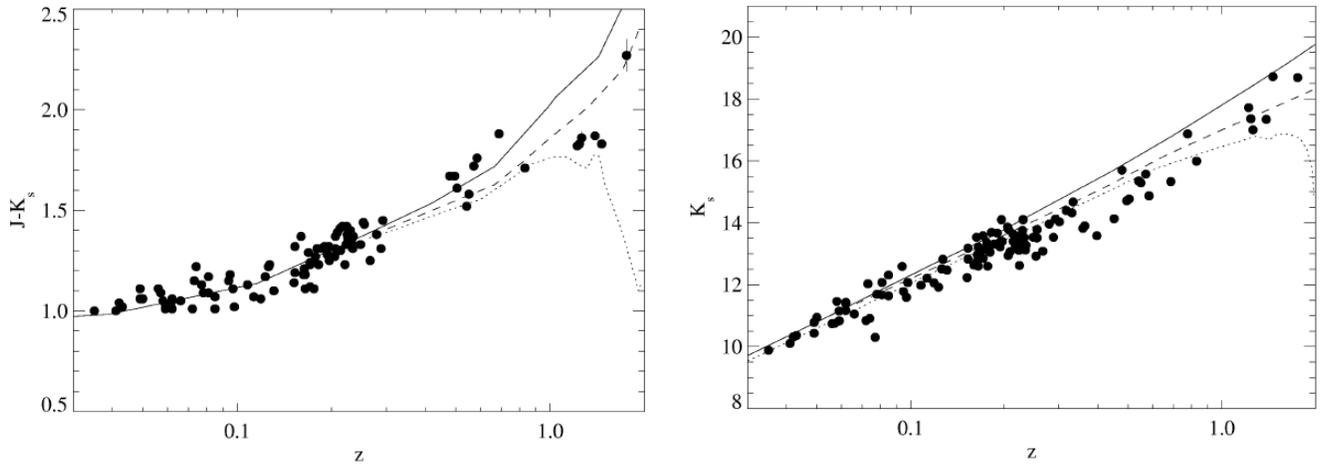

Figure 8. The evolution of stars in BCGs. The J-$K_s$ color (left) and $K_s$ magnitude (right) of the BCG in LH146 compared to lower redshift BCGs from Collins et al. (2009) and Stott et al. (2008). Typical J-$K_s$ errors for z < 1 objects are 0.1. The $K_s$ errors are all smaller than the points. The different line styles indicate different models of the evolution: no evolution (solid), passive evolution from $z_f = 5$ (dashed) or 2 (dotted).

18